\documentclass[9pt,conference,fleqn]{IEEEtran}
\usepackage{amssymb,amsthm,amsmath,array}
\usepackage{graphicx}
\usepackage[caption=false,font=footnotesize]{subfig}
\usepackage{xspace}
\usepackage[sort&compress, numbers]{natbib}
\usepackage{stmaryrd}
\usepackage{xcolor}
\usepackage{mathtools}
\usepackage{float}
\usepackage{textcomp}
\usepackage{nccmath}
\DeclareMathOperator*{\argmin}{\arg\!\min}
\usepackage{nccmath}

\begin{document}
\title{Astronomical high-contrast imaging of circumstellar disks: \texttt{MUSTARD} inverse-problem versus PCA-based methods}
\author{\IEEEauthorblockN{
       S.~Juillard\IEEEauthorrefmark{1}, 
        V.~Christiaens\IEEEauthorrefmark{1}, and 
         O.~Absil\IEEEauthorrefmark{1}\textcolor{red}
    }
    \IEEEauthorblockA{
        \IEEEauthorrefmark{1} STAR Institute, Universit\'e de Li\`ege, All\'ee du Six Ao\^ut 19c, 4000 Li\`ege, Belgium}
}

\maketitle

\begin{abstract}
    Recent observations have shown that protoplanetary disks around young stars can embed a wide variety of features. Raw disk images produced by high-contrast imaging instruments are corrupted by slowly varying residual stellar light in the form of quasi-static speckles. Hence, image processing is required to remove speckles from images and to recover circumstellar signals. Current algorithms that rely on the mainstream angular differential imaging (ADI) observing technique are however limited by geometrical biases, and therefore face a major challenge to reliably infer the morphology of extended disk features. In the last two years, four algorithms have been developed for this task, with three of them based on inverse problem (IP) approaches: \texttt{REXPACO}, \texttt{MAYONNAISE}, and \texttt{MUSTARD}. In this presentation, we will (i) present the new \texttt{MUSTARD} algorithm and (ii) discuss the advantages of IP compared to others methods based on systematic tests.
\end{abstract}

\section{Introduction}
%Please follow the outlined formatting when submitting your manuscript to the International Symposium on Computational Sensing (ISCS). This section provides guidelines for abstract submissions for all types of presentations in ISCS 2023, \textit{i.e.,} platform, poster, and show-and-tell presentations. Specific requirements for show-and-tell abstracts are highlighted in Sec.~\ref{sec:show-and-tell}.

Protoplanetary disks are the known birthplace of planets. Recent observations with ALMA at sub-mm wavelengths and with the latest generation of high-contrast imagers at near-IR wavelengths have revealed a wide diversity of structures in these disks. It is yet unclear how all these structures are connected to embedded planets, considering that protoplanets have been confirmed in only a single system so far \citep[PDS 70,][]{muller, PDS70info, periodPlanetc}. 

High-contrast imaging (HCI) is a particularly challenging field as its primary goal is to enable the detection of faint objects (e.g., exoplanets or circumstellar disks) near very bright stars, which are typically $10^3$ to $10^{10}$ times brighter than the faint objects of interest. The combination of advanced instrumentation with a relevant association of observing strategies and post-processing techniques is required to reveal the faint circumstellar signals hiding in the vicinity of the star. The system's response to an on-axis point source is called point spread function (PSF). The PSF of star, though a high-contrast imager, is corrupted by instrumental aberrations that are slowly varying over time (quasi-static speckles). The majority of the post-processing methods are based on reconstructing PSF models to subtract the stellar contribution from the images.

State-of-the-art algorithms to process ADI observations are focused on exoplanet detection. Only four algorithms have been designed specifically to search for extended sources. Three of them are based on inverse problem approaches \citep[\texttt{REXPACO}, \texttt{MAYONNAISE}, \texttt{MUSTARD},][]{REXPACO, mayo, Sandrine}, and the last one is an iterative algorithm based on Principal Component Analysis \citep{mayo, StapperGinski22}. This raises the question of what are the differences between the three implementations of the inverse problem, and what advantages they have to offer compared to PCA-based methods.

\noindent\textbf{Notations}: 
All the equations presented here share the same notation: %to represent the ADI sequences: 
$Y\in \mathcal{R}^{N,n,n}$ is the ADI cube, with $N$ the number of frame, and where each frame is a square image of size $n\times n$ pixels. Each frame $Y_k$ is associated with the parallactic angle $\theta_k$. The image rotation operation by an angle $\theta_k$ is denoted $R_{\theta_k}$. $S\in \mathcal{R}^{N,n,n}$ is the cube of stellar PSFs with $S_k\in \mathcal{R}^{n,n}$ the stellar PSF in the $k^{\rm th}$ frame and $s\in \mathcal{R}^{n,n}$ the average stellar PSF.
$d\in \mathcal{R}^{n,n}$ denotes the estimated circumstellar signal.
The PCA approximation of the cube is denoted $H_{q}(Y)$, where $q$ is the number of principal components that are kept from the singular value decomposition. This operator provides an estimation of the stellar PSFs $H^{q}(Y) = \Bar{S}$.

\begin{figure}[!t]
    \centering
    \includegraphics[width=\linewidth]{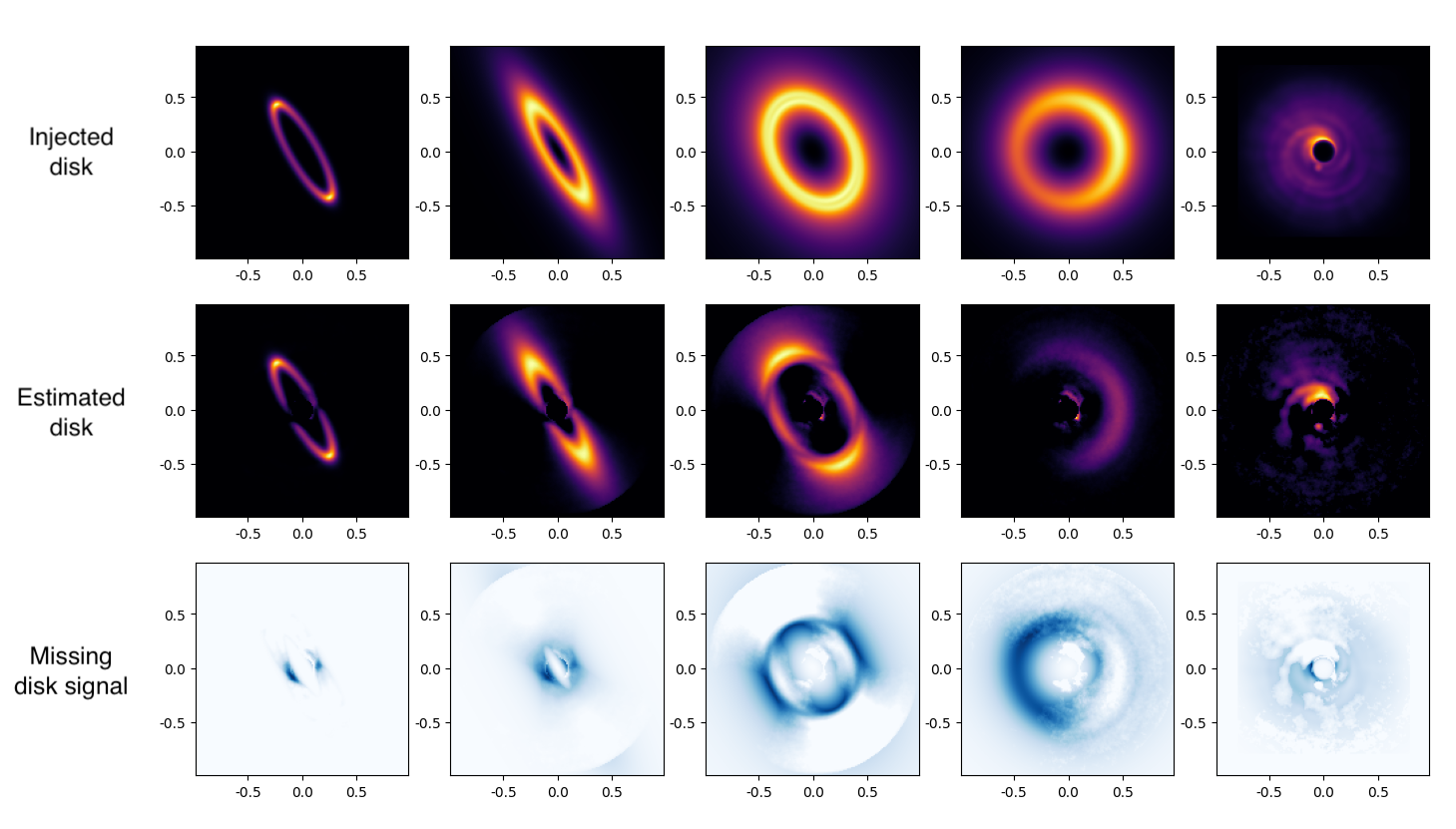}
    \caption{Disk estimations with iterative PCA for a contrast of $10^{-3}$ used for injection in the ADI cube. Comparison between ground truth \textbf{(top)},  iterative PCA estimation \textbf{(middle)} and the map of negative residuals $|\Bar{X} - X|_{max=0}$ \textbf{(bottom)} emphasizing the missing disk signals in the estimation. Axis ticks are set in arc seconds. One can note circular missing  signals, which are the signals invariant to rotation. These residuals depend on the shape of the disk and the amount of rotation available.}
    \label{fig:ambig_circular}
\end{figure}

\section{Context}

PCA is currently the most used post-processing technique, despite being known for inducing ambiguous deformations of extended signals such as disks \citep[][see also Fig.~\ref{fig:ambig_circular} for iterative PCA]{valentin19, Milli, Pueyo16}. PCA is based on the calculation of a principal component basis through singular value decomposition of the ADI sequence, after the transformation of the image cube into a 2D (time vs spatial) matrix. The PCA method creates a PSF model by projection of the ADI images onto a chosen number of principal components. Then the model is subtracted from the original images, and the residual images should contain the circumstellar signal \citep{Soummer12}. Even for a wise choice for the number of components, the method is systematically aggressive towards extended signals, %such as protoplanetary disks, 
such that the latter are not well restored. This is largely due to circumstellar signals appearing in the principal components \citep{Milli, valentin19}.

Several algorithms have been developed recently to tackle the retrieval of extended signals. Iterative PCA has been shown to provide better results than classical PCA. Here we will be using the  \texttt{GreeDS} implementation of iterative PCA \citep{mayo}. This method iteratively refines the stellar PSFs estimation by removing the estimated disk signal at each step to prevent it from being captured into the principal components (and hence in the subtracted PSF model):

\begin{align}
    \Bar{S}_{n+1} &= H_{q}(Y - R(\Bar{d}_n)) \\
    \Bar{d}_{n+1} &= R^{-1}(Y-\Bar{S}_{n+1}) 
\end{align} 

However, when working with extended signals, another limitation of PCA-based methods is the poor handling of the circularly symmetric component of the signal \citep{REXPACO}. This limitation is inherent to the ADI strategy: circularly symmetric components are invariant to field rotation, and will be identified as common to all frames. Examples of deformations caused by signal invariant to the rotation are shown in Fig.~\ref{fig:ambig_circular}. Hence, one purpose of using an inverse problem approach over a blind estimation like PCA is to provide a better interpretation of the data by adding the lacking information through regularization.

\begin{figure}[!bt]
    \centering
    \includegraphics[width=0.7\linewidth]{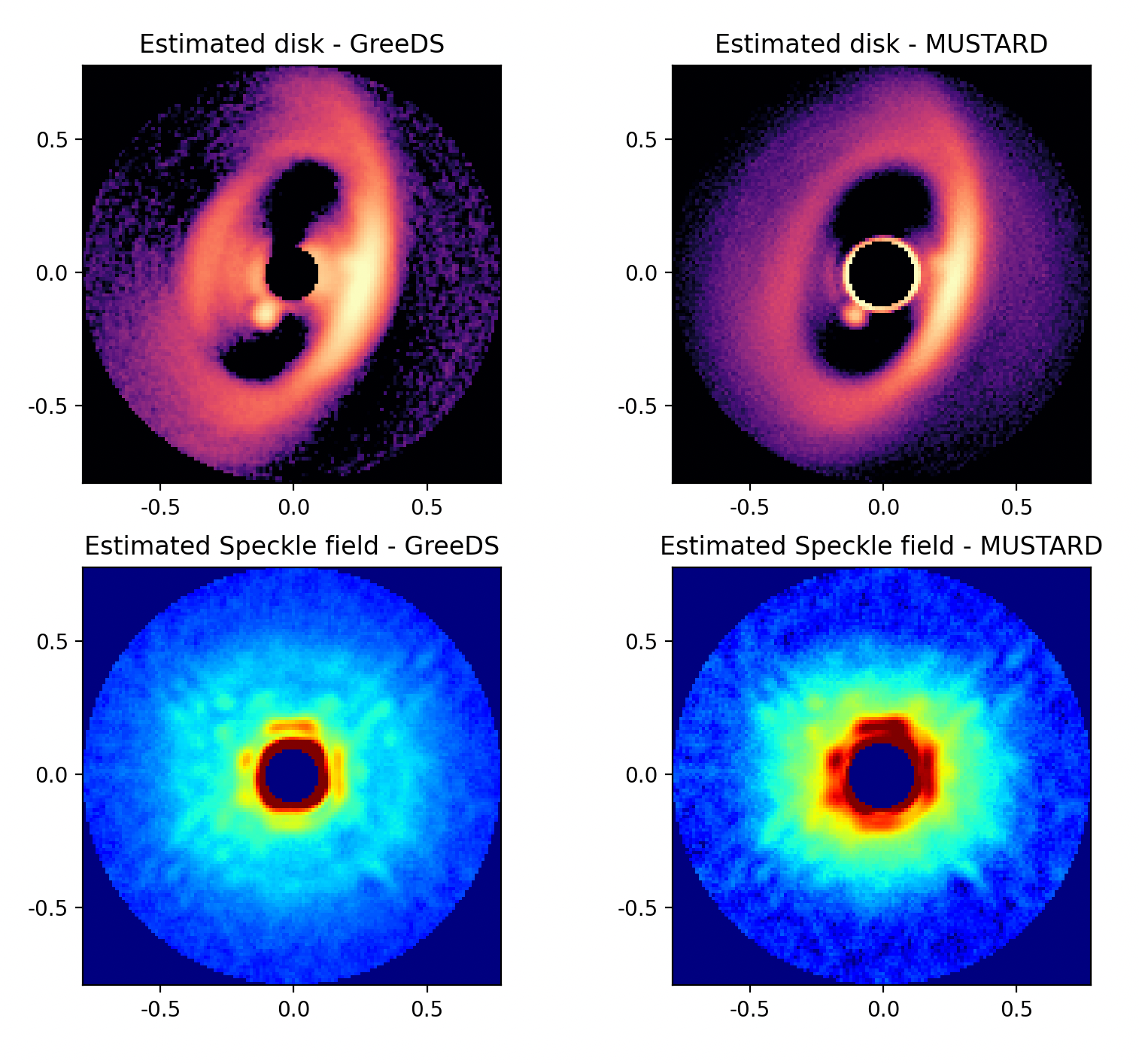}
    \caption{Images of a VLT/SPHERE dataset on PDS-70 (program 1100.C-0481, 2018) obtained using \texttt{GreeDS} (left) and \texttt{MUSTARD} (right), respectively. \textbf{(top)} Estimated circumstellar signal. \textbf{(bottom)} Mean estimated stellar PSF. The color map is arbitrary and in logarithmic scale.}
    \label{fig:PDS}
\end{figure}

\section{MUSTARD approach}

\subsection{Data-attachment term}

%\subsubsection*{\underline{1- Data-attachment term}}

We model the stellar PSF as static throughout the observation and the circumstellar signal as rotating. Both the speckle field and disk signal are considered to be positive signals. \\

\begin{align} 
    Y_k &= s + R_{\theta_k}(d)
    \tag{MUSTARD model}
    \label{equ:y_k}
\end{align}

\begin{equation} 
    \Bar{d} = \argmin_{s\in \mathbb{R}^{+ M,M}, d\in \mathbb{R}^{+ M,M}}(\sum^{M\times M}_{pixels=0}\sum^{N}_{k=0} |Y_k - (s + \mathcal{R}_{\theta_k}{d})|^{2})
    \label{equ:crit}
    \tag{MUSTARD estimator}
\end{equation}

\subsection{Ambiguities and regularization}
%\subsubsection*{\underline{2-Ambiguities and Regularization}}

Our inverse problem is ill-posed due to the overlaps between $d$ and $s$. Indeed, a component invariant to the rotation can be labeled as either static (i.e., as $s$ or rotating (i.e., as $d$). To sort at best the circular ambiguities accordingly, we added a prior on the morphology of the stellar halo. %$M_{\mathrm{reg}}$ is a Gaussian, which represents the way the stellar halo spreads over the frame. Hence, this regularization will penalize certain flux distributions of ambiguous signals according to the shape of the mask.
$M_{\mathrm{reg}}$ is defined as a Gaussian, with a width set to match the spread of the stellar halo. This regularization penalizes the flux distribution of ambiguous signals according to the shape of the mask. While fine-tuning the mask from a dataset to another is expected to improve our results, we rather considered a realistic case scenario of not knowing beforehand the stellar halo profile due to the potential presence of bright disk signals, hence a single average Gaussian profile for all our tests. %This led to satisfying results in most cases. %The mask required fine-tuning although, one typical mask have been used and applied to all test datasets providing satisfying results in most cases. 

\begin{equation}
    R_{\mathrm{mask}} = |M_{\mathrm{reg}}\times d|_{l_1}
    \label{equ:Rl1}
\end{equation}

%\subsubsection*{R2 : smooth regularization}

In addition, we define a smoothness regularization term using the spatial gradient of the estimated circumstellar and stellar signals $d$ and $s$, respectively. It is a common regularization to compensate for noise. With $\Delta$ standing for the Laplacian operator, the regularization term reads:

\begin{equation}
    R_{\rm smooth} = ||\Delta d||^2 + ||\Delta y||^2
    \label{equ:Rsmooth}
\end{equation}

A good balance between the two morphological regularization terms and the data-attachment term is required to provide good results and avoid over-fitting of the mask. Angular diversity is not adapted for extended signal due to signal invariant to the rotation. So far, \texttt{MUSTARD} is the only approach to address this limitation.

\begin{figure}[!t]
    \centering
    \includegraphics[width=1\linewidth]{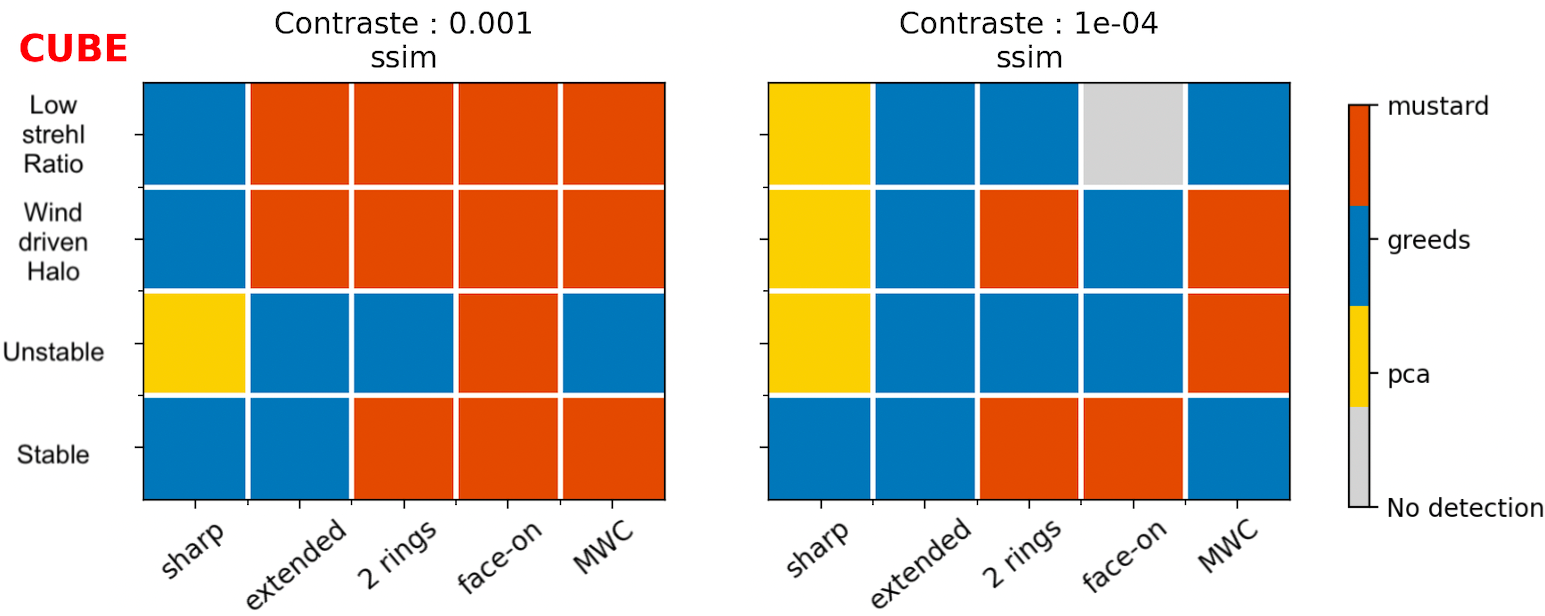}
    \caption{Result of systematic test performed to compare PCA, Iterative-PCA (\texttt{GreeDS}) and Inverse problem (\texttt{MUSTARD}). Each cell represent one simulation. We simulated 5 different disk morphologies \textbf{(x-axis)} and injected them in 4 different star observations obtained with the VLT/SPHERE/IRDIS instrument in various observing conditions \textbf{(y-axis)}. The injection was performed at two different contrast levels: $10^{-3}$ \textbf{(left)} and $10^{-4}$ \textbf{(right)}. The color of the cell represents which algorithm performs the best according to the SSIM metric. Gray cells mean that no algorithm detected the injected disk (upon visual inspection). %Non-detections are assessed by eyes. White letters on a cell indicate when one or two algorithms did not detect the disk (p$\longrightarrow$PCA, g$\longrightarrow$\texttt{GreeDS)}, m$\longrightarrow$\texttt{MUSTARD)}
    }
    \label{fig:results}
\end{figure}

\section{Results}

We compare the processing of a dataset of PDS 70 with \texttt{GreeDS} and \texttt{MUSTARD} in Fig.~\ref{fig:PDS}. In the \texttt{GreeDS} estimation \textbf{(left)}, we can visually identify some stellar halo signal that is not properly handled hence contaminating the disk estimate, and vice-versa. In the \texttt{MUSTARD} estimation \textbf{(right)}, the erroneous assignment of flux invariant to rotation is corrected according to the regularization prior.

We injected five synthetic disks in four empty datasets tracing different observing conditions and total field rotation (stable, 80°; unstable, 91°; presence of wind-driven halo, 85°; low Strehl ratio, 26°; respectively), and at two contrast levels (1e-3 and 1e-4). We then tested the recovery of the injected disks using (non-iterative) PCA, \texttt{GreeDS}, and \texttt{MUSTARD}. We defined the best reconstruction based Structural similarity index measure (SSIM)\citep{SSIM}. This metric aims to capture the disparity of contrast, luminance, and structure on various windows of the images. Results are summarized in Fig.~\ref{fig:results}. 

We observe that the inverse problem approach is able to provide good results in the most favorable cases (bright disk, stable datasets) with an optimal reconstruction error and a better flux estimation. \texttt{MUSTARD} is especially better than PCA-based solutions for disks that contain more flux invariant to the rotation (i.e., the more face-on disks from our samples). On the other hand, iterative-PCA is more robust and easier to parametrize; it is better than a simple PCA in most cases, and it is better than \texttt{MUSTARD} when the speckle field is unstable and the contrast is low.

One of the main limitations to our \texttt{MUSTARD} inverse problem approach is related to our limited knowledge of the dynamics and morphology of the speckle field, and of some known nuisance phenomena that we are not able to describe mathematically, which are limiting the quality of the model.

\bibliographystyle{IEEEtran}
\bibliography{references}
\end{document}